\documentclass[12pt]{article}
\usepackage{sbc-template}
\usepackage{graphicx,url}
\usepackage{amsmath}
\usepackage{color}

\usepackage[brazil]{babel}   
\usepackage[utf8]{inputenc}

\title{Eletronic Health Records using Blockchain Technology\footnote{This research is part of the INCT of the Future Internet for Smart Cities funded by CNPq, proc. 465446/2014-0, CAPES proc. 88887.136422/2017-00, and FAPESP, proc. 2014/50937-1.}}

\author{Arlindo F. da Conceição\inst{1}, Flavio S. Correa da Silva\inst{2}, Vladimir Rocha\inst{3},\\ Angela Locoro\inst{4} and João Marcos M. Barguil\inst{2}}
\address{
  Universidade Federal de São Paulo (Unifesp)
  \nextinstitute
  Universidade de São Paulo (USP)
  \nextinstitute
  Universidade Federal do ABC (UFABC)
  \nextinstitute
  University of Milano-Bicocca (UNIMIB)
  \email{arlindo.conceicao@unifesp.br, \{fcs, jbarguil\}@ime.usp.br, }
  \email{vladimir.rocha@ufabc.edu.br, angela.locoro@unimib.it}
}

\begin{document} 

\maketitle

\begin{abstract}
Data privacy refers to ensuring that users keep control over access to information, whereas data accessibility refers to ensuring that information access is unconstrained. Conflicts between privacy and accessibility of data are natural to occur, and healthcare is a domain in which they are particularly relevant. 
In the present article, we discuss how blockchain technology, and smart contracts, could help in some typical scenarios related to data access, data management and data interoperability for the specific healthcare domain.
We then propose the implementation of a large-scale information architecture to access Electronic Health Records (EHRs) based on Smart Contracts as information mediators.
Our main contribution is the framing of data privacy and accessibility issues in healthcare and the proposal of an integrated blockchain based architecture.
\end{abstract}
     
\section{Introduction}

Data Health interoperability is a problem that remains open until now. The main question is how to provide open access to sensible data (health data), preserving personal data privacy, anonymity, and avoiding data misusage.

Blockchain technology~\cite{nakamoto2008bitcoin} and smart contracts seem to provide an interesting and innovative way to keep references to Electronic Health Records (EHRs). 
By using this technology, the patients could keep a better control of their own data and health professionals and institutions, such as hospitals, could have access to patients data owned by others institutions. In summary, blockchain has the potential to improve EHR solutions providing privacy and interoperability.

This article explores how blockchain and smart contracts can be applied to improve EHR. To do this, first, we discuss some nontechnical aspects that make health data sensible. Then, we propose an architecture that is able to improve current EHR systems. Our goal is to provide secure access to patient's data, avoiding a third party access it without permission; only the patient and health professionals or institutions should be involved.

In order to present and discuss these issues more effectively, we introduce one scenario that is particularly relevant:

\begin{itemize}
\item \textbf{Basic scenario. }Patient \textit{X} needs some medical data while having a medical appointment at institution \textit{Z}, but the data was created during a previous appointment between \textit{X} and \textit{Y}, and the data is kept by professional/institution \textit{Y}. In this scenario, \textit{X} is usually a regular person. \textit{Y} and \textit{Z} are different health professional or health institutions.
\end{itemize}

The scenario above is very common but creates several technical and nontechnical problems. The first question is how to find and to provide accessibility to \textit{X}'s health data? This implies the existence of a discovery service and the interoperability between \textit{Y} and \textit{Z}. The second question is what privacy restrictions must be observed? For example: \textit{i}) privacy of data, \textit{ii}) control of who can access each piece of data, \textit{iii}) trust in health institutions, etc. And, even if all data are human readable, e.g. using PDF format, the data format standardization is an additional challenge.

Data privacy refers to ensuring that users keep control over access to personal information \cite{pavlou2011state}, whereas data accessibility refers to ensuring that information access is unconstrained \cite{culnan1984dimensions}. Conflicts between privacy and accessibility are natural to occur, and healthcare is a domain in which they are particularly relevant.

Most of the times, patients need care from different physicians and different institutions that should, in their turn, be able to access other physicians' clinical notes and health records~\cite{reti_improving_2010}. Institutional health records are not always open to other institutions, and institutional or personal health records are not always interoperable~\cite{detmer_integrated_2008}. These issues are well known in healthcare, and has to do with a fundamental problem in medicine, which is known as \emph{care coordination} (see for example~\cite{klein_use_2015}). 

In  other scenario, if the medical condition presented by the patient prevents her from being capable of granting permission to her records (e.g. because the patient is unconscious), we can consider a ``break the glass'' mechanism, where a health professional is granted permission to access her records, given explicit account of the patient's condition and presentation of required credentials of the professional, as well as publicly available information stating that the patient has explicitly declared that she would agree with this permission under clearly specified conditions.

For the benefit of the patient, the number of mediating institutions used to implement these scenarios should be minimal, to reduce information curating costs as well as requirements to trust in a reduced number of third-party institutions. 

A solution based on blockchain can enable large-scale accessibility, keeping data privacy, reducing curating and mediating costs, as well as providing \textit{trustless trust} in information systems.
In the following sections, we discuss blockchain related work, data accessibility, and privacy issues in the light of the implementation and maintenance of a global EHR based on a blockchain decentralized architecture. We further outline some underlying principles of this system, in terms of ethical issues and openness, and we provide an architectural sketch of it, which may encompass all these properties and principles.

\section{Blockchain: distributed and secure coordination}
The most popular usage of blockchain technologies was in the cryptocurrencies, such as Bitcoin \cite{nakamoto2008bitcoin}. More recently, they have been proposed as means to implement other kinds of decentralized applications \cite{ferrer2016blockchain,lazarovich2015invisible,lewenberg2015inclusive,norberhuis2015multichain,petersonblockchain}.

Blockchain technology is grounded on the notion of a \textit{distributed ledg\-er}, which acts like a database containing information about the history of transactions involving those agents. It is constantly audited by groups of agents (selected according to different policies, depending on the application domain).
The result of each auditing is stored in a \textit{block} and broadcast to the network. Blocks are sequentially appended to the ledger, forming a cryptographically-linked \textit{chain}.
Attempts to tamper the blocks or to alter their order can be easily detected. 

The whole community may accept or reject the reliability of any block,
according to a predefined set of rules.
If an agent receives several valid additions to their local copy of the ledger, they always choose the longest chain of valid blocks (or the earliest one, if they have the same length), ignoring other conflicting and less relevant chains.
This conceptually simple procedure ensures that \textit{consensus} is eventually reached, even in scenarios where propagation is slow due to high network latency.

Similarly, ill-intentioned nodes may try to insert malicious entries in the ledger, but the community will simply reject their blocks and ignore their chain, effectively forcing them to abide by the rules.

If auditing is approved by the community, then the ledger -- possibly containing recent, previously unverified transactions -- is replicated across the agents. Otherwise, the largest accepted portion of the ledger is replicated with information about dissonances and corresponding actions to be taken -- whose effects are then registered as new transactions to be audited in future rounds of verifications.

A blockchain based solution can, therefore, be envisaged to ensure accessibility of information in any large-scale system. It can be most appropriate for the distribution of health records across a network of healthcare agents, provided that solutions are given for latency and storage requirements related to it, given that (1) peers are required to store copies of the ledger of interactions and (2) transactions and blockchains \textit{per se} must be disseminated across the network of peers.

Ethereum \cite{wood2014ethereum} is a blockchain-based platform for fully decentralized applications. It is based on the notion of \textit{smart contracts}, which are procedures that determine sequences of actions in order for peers to interact with each other. Smart contracts can be used to implement agents that are relevant to manage information. For example, smart contracts can contain rules to discipline access to the contents of encrypted health information. In this way, smart contracts could allow implementing a privacy layer in a distributed information system.

The architecture proposed in this article is inspired by the current version of Ethereum platform and in the idea of automatic health data workflow using smart contracts.

\section{Related work}
In the recent years, several authors explored the idea of using intelligent agents in healthcare context to provide interoperability~\cite{Isern2016, CORTES2015, Wimmer2014}. More recently, some authors explored blockchain technology~\cite{shrier2016office, liu2016medical}, most of them have a theoretical approach -- as we did, proposing strategies to improve mobility and security in EHRs using distributed ledgers. As expected, we found only a few references discussing real implementations of EHRs over blockchain~\cite{azaria2016medrec, wood2016blockchain}, since blockchain is a relatively new technological development.

The \emph{Cyph MD}\footnote{ Information about Cyph MD can be found in \url{http://www.startupdaily.net/2016/08/cyph-md-blockchain-healthcare}. Accessed on February 2018.} is a proposal of an Australian startup to manage health records using Ethereum framework. We could not find further details about the product, we suppose that it is in the very first stage of development. 

The GemOS~\cite{wood2016blockchain} is an Operational System (OS) that claims to provide security based in blockchain as an OS feature. It provides a common ledger and allows secure key management and verification\footnote{See further details in \url{https://gem.co/health}. Accessed on March 2018.}. Two applications of GemOS are supply chains and EHRs. However, we could not find details of practical deployment of GemOS.

Azaria et al.~\cite{azaria2016medrec, ekblaw2016case} provide a proof-of-concept that uses block\-chain as a mediator to health information. 
The prototype, called MedRec, brings medical researchers and healthcare stakeholders to ``mine'' in the network and, as a reward for mining, it releases access to aggregated and anonymized medical data. The authors argue that a sustainable and securing peer-to-peer network can be built just by supplying big data, in order to empower researchers while engaging patients and providers. The platform MedRec, however, was validated only for medical records and needed to be extended for more critical health data and complex scenarios~\cite{halamka2017potential}.

\section{Requirements for an open and global EHR system}

In this section, we discuss some non-trivial requirements of EHR systems and how blockchain and smart contracts could help in met these requirements.

\subsection{Data accessibility and management}

In the scenarios considered in this article, accessibility to health records is a fundamental challenge. Orthodox solutions would be based on formal and social acceptance of one (or a network of) mediating institution(s), who would be in charge of storing and curating health records, as well as controlling access to them according to established rules and contracts.

Two relevant issues related to the establishment of such mediators are:

\begin{enumerate}
\item \textbf{Cost} related issues, as -- even if the mediators are nonprofit institutions -- the maintenance of a reliable infrastructure to store health records in large scale and control access to them in appropriate manner demands significant resources.

\item \textbf{Trust} related issues, as such mediators must be regarded as reliable and trustworthy by all stakeholders, lest they will not be accepted as information keepers.
\end{enumerate}

Blockchain technologies have enabled the alternative of having a decentralized network of mediators to store and manage information. In this technology, the higher and less structured the network of mediators is, the more reliable the \textit{collective behavior} of these mediators can be considered, without the necessity to trust any specific mediator \cite{fadhil2016bitcoin}. In this way, blockchain technologies introduce the notion of \textit{trustless trust}, as users of an information system can trust the system as a whole without the need to acknowledge or trust any specific peer.

\subsection{Privacy levels and anonymization}

Personal health data is a sensible data and must be kept in secret. An EHR system has to implement privacy policies in order to ensure that only the own patient and the healthcare agents, who have explicitly granted permission by the patient, have access to personal health records.

However, aggregate and de-identified information could and should be accessible to certified institutions in charge of the management of public healthcare, e.g. to monitor and prevent the spread of epidemics. 
A \textit{commitment and encryption} layer must be added to the solution to cater for different privacy issues. 

Exceptionally, access to personal data should also be considered for those situations in which a patient is not responsive (e.g. because she is unconscious). These exceptions require the participation of mediating third party institutions or other care providers, which must register, store and manage the exceptions (declared by individuals) as well as credentials, and the situations that demand exceptional access to private information.

In addition, anonymized aggregate data must be available to authorities in charge of public healthcare. In this context, data anonymization is a type of information sanitization whose intent is privacy protection. It is the process of either encrypting or removing personally identifiable information from data sets so that the people to whom the data refers remain anonymous. But, in some cases, anonymization still has a risk of negative social consequences \cite{Taylor2016}. Corresponding risks must be clearly presented to patients prior to a subscription to any information management system.

Thus, privacy has to be organized in sub-layers, such that access to each layer is controlled by different sets of rules.  For example, when an electronic health record is being built, carefully designed templates can be used in order to ensure that pieces of information are stored in appropriate sub-layers. Anonymised aggregate data can be in one layer, and individually identified data can be in a separate layer, which can only be disclosed upon direct patient consent.

Smart contracts could be a tool to the process of data sanitation, automatizing the organization of data in different levels of privacy and details. In addition, consensus algorithms over blockchain could help to create the ``break the glass" solutions without the need of a third-party mediator.

\subsection{Ethical issues}

Health data management provides opportunities to improve the quality of life of citizens. However, these opportunities create some ethical challenges~\cite{Mann2016}.
Based on the analysis of the possibilities for accessing healthcare related data, Floridi and Taddeo~\cite{Floridi2016} outlined three axes of ethical questions related to information: the ethics of \emph{data}, the ethics of \emph{algorithms}, and the ethics of \emph{practices}. 

\begin{enumerate}
\item \textit{Ethics of data} relates to data ownership, transparency, and privacy. It has the power to clarify issues related to the blurred line which divides what should be open data and what should not. 

\item \textit{Ethics of algorithms} relates to the established behavior of a program or intelligent agent in order to avoid, for example, wrong decisions or unintended exposition of information. 

\item \textit{Ethics of practices} relates to how information is used, according to professional ethical tenets which, for example, establish respect for Human Rights and the need to avoid negative outcomes such as discrimination or unauthorized or malicious use of information.

\end{enumerate}

The design of a global EHR using blockchain must -- at least try to -- deal with open questions in each of the three axes. Some questions are heavily influenced by cultural issues and unforeseeable circumstances. Regarding ethical dilemmas, we try to depict some of them for the specific healthcare domain and EHR applications:

\begin{itemize}
\item \textbf{Data ethical dilemmas}: The respect for patient autonomy is almost uncontroversial. There are few situations where patients may not have the right to decide which procedures they wish to undergo, or which data they wish to have exposed.
Sometimes there can be a public interest in individual data, which can raise an issue about over-ruling individual patient rights for the benefit of a social group. For example, if we consider the regional control of infectious diseases such as tuberculosis, it can be important to track the incidence and prevalence of cases at the individual level, regardless of personal consent provision by patients. Also in emergency cases, a patient may not be able to make autonomous decisions and medical professionals should be able to request over-ruling of individual rights to safeguard the life of patients.

\item \textbf{Algorithmic ethical dilemmas}: Imagine a scenario where a patient urgently needs a bone marrow donation and a global EHR management system is able to identify a potential donor, but the system cannot reveal the identity of that donor. The dilemma presented here rests in the decision between trying to save a life and respecting individual rights. Additionally, imagine that, in this same scenario, the system can be either of a proprietary source or open source. The opaqueness of procedures employed to solve the dilemma clearly is an issue to be solved. It should be observed, nevertheless, that the issue here is not so much about the access permission and the power to modify source code -- as usual, we understand that \textit{open source} software is about -- indeed it is about the scrutability and the verifiability of those procedures.

\item \textbf{Practices related ethical dilemmas}: During a medical appointment, both patients and health professionals can act wrongly, in a non-sincere or unprofessional way. For example,  a patient can give a wrong information during a medical appointment because of health illiteracy, misunderstanding or privacy reasons; a physician can omit the side effects of a medicine because he has an interest in promoting a particular brand of medicine.
\end{itemize}

A privacy management system, auditable and based on reputation, could have a positive impact in these kinds of malpractices. For example, a doctor may think twice before undertaking a procedure that can be audited and cause controversy. As a drawback, if malpractice may leave a trace, some patients may prefer to self-medicate, which may have negative consequences for the quality of care.

\subsection{Openness}

To build a fully decentralized system requires trust on several components, such as third-party institutions, agents and the information network itself. In this section, we argue that openness can be fundamental to reach the desired level of trust in the system.

Openness can be defined as an emphasis on transparency and free, unrestricted access to knowledge and information, as well as collaborative or cooperative management and decision-making rather than a central authority. Openness can be said to be the opposite of secrecy.

At least four different aspects of openness are useful in a global EHR and are also present in blockchain technology: open or free software, open standards, open data, and open innovation. We expose each of these aspects in the next paragraphs:

\begin{itemize}

\item Regardless of the system architecture adopted in a global EHR, this system can benefit from the adoption of open source software to ensure that information is processed in trustworthy ways. Open source in this scenario is useful not so much as a means to have community-based software development, rather as a means to ensure auditability of code, for which open source is a necessary but not sufficient condition. In fact, additional discipline for software development must be a place to ensure auditability~\cite{Ryoo2014}. Furthermore, in open source solutions, accreditation of software code must be provided, so that institutions worldwide can ensure that the software tools they use are the right ones to be used. The usage of open source code can also facilitate the finding and correction of errors in the system. The adoption of free software licenses facilitates the sustainability and the long-term collective improvement of the system and, of course, can reduce costs.
Interestingly, whereas reducing costs seems to be the major benefit of free software, also the quality of the final product is a nonnegligible and important benefit.

\item Similar arguments about auditability and accreditation also apply to the adoption of open standards to encode and exchange information. In addition, to use open standards is strictly important to system integration.

\item Open data is not required in the specific system proposed, as we have privacy-sensitive data flowing across a network of peers. 
However, open data is desirable as a source of information to understand network dynamics. For example, it may be used to add a social aspect in the system, i.e., enabling agents to allow or deny information access to an institution depending on its reputation or its public perceived trust. In other words, although open data is not a central feature in the system, it may, however, be a powerful tool for improving its future releases.

\item Open innovation is more related to digital culture (see \url{http://digitalprinciples.org}) than to software technology. It implies the creation and involvement of a community of developers and users to solve social problems. It means to encourage a free flow of ideas and the free sharing of knowledge and solutions. It also means to recognize the usage of open data, open standards, and free software as an investment in a public good.

\end{itemize}

Nowadays openness, in all of its aspects (open data, open standards, free software and open innovation) is a requirement to build trusted software solutions. Both EHR and blockchain applications are particularly interesting cases, as it may unfold the different impacts that each of these concepts can have and how they should be considered in order to build effective solutions for complex problems such as those encountered in the healthcare domain.

\section{A general architecture for a global scale EHR based in blockchain} 
\label{sec:archi}
 
Given the above considerations about data accessibility, privacy, ethical issues, and openness importance, we introduce an architecture of  EHR, based in blockchain and smart contracts, that could make health records interoperability possible and safe on a global scale.

Our architecture, illustrated in Figure~\ref{arch}, has the following components:

\begin{figure*}[hbt]
\centering
\includegraphics*[width=\textwidth]{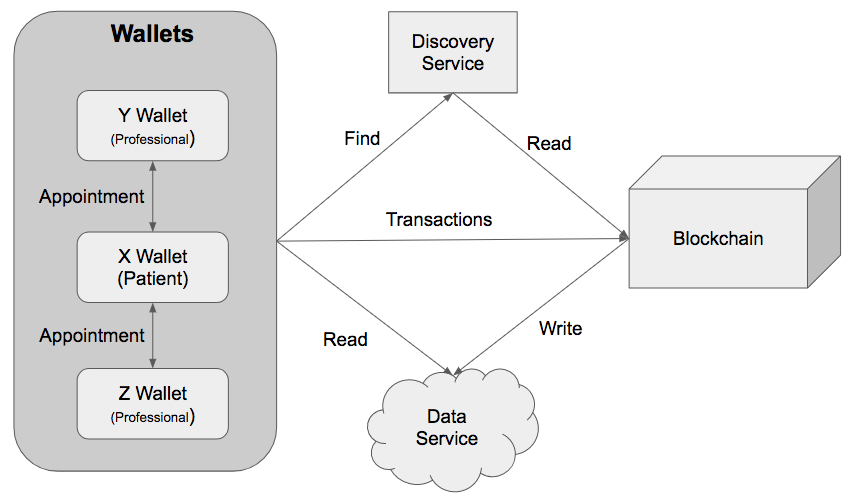}
\caption{Proposed architecture.}
\label{arch}
\end{figure*}

\begin{itemize}
\item \textbf{Blockchain}. A distributed ledger, capable of to execute smart contracts. This component is responsible to record references to health transactions, such as health appointments, clinical exams, prescribed medication, etc. The format of each transaction is described in Section~\ref{transactions}; in summary, in a cryptocurrency system, a block contains currency transactions. In a  privacy layer for electronic health records, a block can contain pointers to health information. For example, when a patient $X$ is seen by a doctor in the hospital $Y$, a transaction is appended to the ledger saying that $Y$ have access to that information of $X$.

\item \textbf{Data Service}. A data storage service necessary to keep the health records. 
In this proposal, it is a cloud file system, where each file is owned by \textit{X} and can be read by \textit{Y}\footnote{It is quite important to note that in this approach the data is owned by the user. This approach is not followed by traditional EHRs, where the data is owned also by the health institution. We put under attention here the complex question of health data property. We will argue in the future that the patient should be remunerated by the usage of its information.}. This solution can be implemented, for example, by using popular cloud services in the market, such as Google Drive, Megadrive, and Dropbox. 
To be used in our architecture, the data service must provide: cloud access, file access control and interfaces to add and remove reading access to the files.
\item \textbf{Wallets}. An electronic wallet is responsible to store the users private and public cryptographic keys. The public key is the user identification in the solution. 
The email and password used to access the data service are also kept by the wallet.
The wallet is the basic interface and method of access to the system.
\item \textbf{Discovery Service}. This nonmandatory and auxiliary system is used to accelerate information search. It is an index to the information stored in the blockchain. For example, given a patient \textit{X}, identified by his public key $X^+$, the discovery service offers a list of transactions in the blockchain owned by \textit{X}. It should also offer an interface to find \textit{X} and \textit{Y} given a pointer to a data file. This service could be implemented using a NoSQL base that keeps a view (with eventual consistency) of the blockchain. There is no security issue related to this component because it only read the blockchain and the queries results can be easily verified by the local copies of the blockchain. In addition, the Discovery Service can evolve to offer basic services of a search for professionals.
\end{itemize}

The architecture separates the transaction control (made using the blockchain ledger) from the data storage (Data Service). We could imagine a solution where all data is stored in the ledger, but, for performance reasons, it is not feasible. 

One characteristic of this architecture is to delegate the data management to the users. The patient owns the data and can delete or restrict access to it whenever he or she wants.

The core of the architecture is the set of smart contracts. They are stored in the ledgers and are responsible for:

\begin{itemize}
\item Store a new transaction in the ledger.
\item Receive and process access requests.
\item Register all data access granted.
\end{itemize}

\subsection{Transactions}\label{transactions}
A health transaction is the basic unit of information manipulated by the system. The following types of transactions are defined:
\begin{itemize}
\item \textbf{New Record}, this transaction creates a new record in the ledger. It contains the following fields: \texttt{timestamp}, $X^+$, $Y^+$, metadata about the content, public data about the content, \texttt{link}, \texttt{hash(data)}. \\

Where: the metadata and the public data are optional fields; the \texttt{link} is the path for the electronic health data, stored in the cloud, in a specific format and encoded in such a way that only \textit{X} or \textit{Y} could read it; for example, the sensible parts of the file can be encoded using a key $K$, but $X^{+}(K)$ and $Y^{+}(K)$ is contained in the file. The field \texttt{hash(data)} allows to verify the fidelity of the data content, it is essential to verify that the content was not modified.

\item \textbf{Request Access}, this is created by \textit{Z} to request access to a content owned by \textit{X}. It contains: \texttt{timestamp}, $Z^+$, $X^+$, $Z^-(X^+($link$))$. The term $Z^-(X^+($link$))$ can be used to confirm the identity of Z. Note that only X can grant access to the data because the link is encoded with $X^+$.

\begin{itemize}
\item \textbf{Access Granted}, this is created by \textit{X} in response to a \textbf{Request Access}. It contains: \texttt{timestamp}, $X^+$, $Z^+$, \texttt{link}. Note that link will be a copy of the original value, with permissions to \textit{Z}
\end{itemize}

\item \textbf{Notification}, this is a special transaction created by \textit{Y} to report public health issues, such as cases of malaria, dengue, and tuberculosis. It contains: \texttt{timestamp}, $Y^+$, period of time, and $Y^-($link$)$. A low level of data anonymization is obtained by aggregating notifications in only one transaction; to implement this, a period of time must be provided.
\end{itemize}

\subsection{Data ownership}
In this system, the data ownership is from the patient \textit{X}. It is assumed that each user configures a cloud file system in the wallet. The user can not delete a transaction in the ledger, but he can delete his own health data, effectively revoking access to it. 

\subsection{Wallets}
The system relies on the mechanism of \textit{personal health wallets} that contains, in a secure manner, the patient's identification to a blockchain. The Wallet does not contains 
A patient is not identified in the blockchain, only a public key is exposed. However,  it is known that an identity can be exposed by investigating other factors~\cite{goldfeder2017}.

\subsection{Smart contracts}

A smart contract is a piece of software that is stored in the blockchain and can be executed in a virtual machine. They are used, for example, to execute a new transaction, to regulate conflicts in transactions, send alarms, etc.

In our architecture, contracts should validate the transaction requests. The contract is, in fact, the software component that, after validation, sends a transaction to be stored in the blockchain. 

We predict the necessity of, at least, the following contracts:
\begin{itemize}
\item \texttt{CreateNewRecord}
\item \texttt{ProcessAccessRequest}
\end{itemize}

Other contracts can be developed to implement elaborated anonymization routines, to remunerate patients by data access granted, etc.

\section{Scenarios of usage}

The blockchain stores the history of interactions between patients and healthcare agents, together with links to the EHRs that contain detailed descriptions of each of the interactions. A hash of the medical record is also stored in the block, thus the fidelity of the content can be easily verified.

Whenever a patient \textit{X} meets a healthcare agent \textit{Z}, the agent should have the appropriate software tool, the Wallet, to retrieve health data from the block\-chain. The professional \textit{Z} can request data from \textit{X} and ask the right to read the data. The user \textit{X} must explicitly authorize the access. In this case, new transactions are created in the blockchain, for each file is created a transaction authorizing \textit{Z} to have access to that file. Note that \textit{Z} can choose, based on the metadata content stored in each transaction, to have access to specific files. For example, \textit{Z} can request access only to the files relevant to the current medical appointment.

In principle, each patient can have his/her own wallet instance, whose behavior can be determined according to rules set and combined by the patient. For example, patients should be able to specify the behavior of the Wallet in case they are not responsive and urgent care must be provided. Additional rules can be created, adding, for example, a procurator.

Another important scenario happens when \textit{Z} requests access directly from \textit{Y}. Again, a new transaction must be registered, giving access to \textit{Z} those documents kept in the data service of \textit{Y}. See that this imply that the content of \textit{X} was copied, so the wallet of \textit{X} must be notified about this kind of transaction.

Moreover, specific anonymized information should be accessible from the database, e.g. to support the control of epidemics and the spread of contagious diseases. In fact, this architecture can be easily extended to support different levels of privacy. One first level could be implemented by reading transaction's metadata. The metadata is used to guide professionals about the content of a transaction, but it can offer also some hint about health conditions. For example, suppose that, by reading the metadata, it is possible to note several requests for exams to verify the diagnosis of dengue. The users and institutions are not identified, but it is a relevant indicator of population health.
Beyond of reading open data in the blockchain, it is simple to extend the architecture to implement new types of transactions to provide a better view of the populational health.

Finally, it is important to note that by using the architecture proposed in this article does not prevent that health institutions continue to use their EHR systems. In the first moment, both systems should be used until all legacy systems could be adapted.

\section{Considerations about implementation}
As we have already mentioned, the ledger component could be implemented using the Ethereum platform and its smart contract solution. The Wallets could have a mobile version implemented, for example, using Android platform. A Wallet should be developed taking into account the risks of exposure of sensitive data and should minimize such possibilities~\cite{dehling2015exploring}.
For data storage, we should implement interfaces for the most popular cloud services. The Google platform offers a programming facility called Google Apps Script (GAS) that helps the integration between wallets and the data repository (Google Drive, in this case).

The simplest implementation of the discovery service is, periodically, analyses the blockchain and update the indexes. The database of this service could be a document-based NoSQL implementation, such as MongoDB. Besides, the discovery service should be based on web services interfaces to provide interoperability. In order to avoid a single point of failure, the discovery service should be replicated and each Wallet should have references for more than one instance of discovery service.

\section{Open questions about Proof of Work and economic ecosystem}

The proof of work in a blockchain provides an indication of the reliability of a network node. In Bitcoin, it is the amount of processing power spent. In our context, the active participation in the network should be inferred by the number of the new transactions recently done. The time that a node is present on the network may be another parameter; the longer is this time, more reliable should be the node.

There are several open questions about the economic potential of the system. First of all, what is the value of personal health data? The system can, for example, reduce the number of clinical exams and improve medical appointments quality. However, who will pay for it? And how much \textit{Z} will pay to have access to a health file? Could these values be dynamic? To answer these questions is important to define rules of the system.

Another question is the adoption rate of the system. At the first moment,  the system bootstrap, a global EHR seems to be practical for patients and liberal professionals. But for hospitals and big health institutions, that already keep control of a large amount of health data, the system can be interpreted as a risk for a well established and profitable business model.

Because in our model users are the owners of medical data, this means that they must have control of the corresponding private keys. This may present a challenge, as users who are not tech-savvy (and even ones who are) may lose their keys, resulting in data loss or, even worse, data compromise. 

To better understand these questions, it is necessary to validate the system with real patients and health professional.

\section{Future Work}
For the future, we plan to implement a functional prototype of the proposed architecture, shown in Section~\ref{sec:archi}. This minimal implementation should include a simple mobile wallet written in Android~\cite{elenkov2014android} and the contracts using Ethereum framework~\cite{wood2014ethereum}. Our goals, besides the architecture validation, is to evaluate empirically the number of transactions supported by the blockchain. Another aspect to be evaluated is the capacity of Ethereum development framework to express complex computer models, which are necessary for the implementation of -- real smart -- contracts in the healthcare domain.

\section{Conclusions}
Nowadays, despite recent advances in IT, few health institutions provide data integration between units. This work exploits the main healthcare scenarios of care coordination and health information management and proposed an architecture for secure data exchange. In our proposal, all data is owned by the patients. And it relies on blockchain technology and well-diffused cloud storage services to obtain security, high availability, fault tolerance and improved trust.

The proposed architecture is flexible in order to cater for the considerations about technical and ethical issues. It does not solve 
-- nor point to solutions for -- 
dilemmas related to data accessibility versus privacy, but it provides appropriate means for the explicit consideration of each of these issues and; in this way, it contributes a step ahead towards their solution. In the future, after to create a reliable health data network, we could delegate to the blockchain the activity of data access grant, opening new frontiers to data health management.

\bibliographystyle{sbc}
\bibliography{sbc-template}

\end{document}